\documentclass[12pt]{article}

\usepackage{newtxtext,newtxmath}

\usepackage{graphicx}

\usepackage[letterpaper,margin=1in]{geometry}
\usepackage{braket}

\renewenvironment{abstract}
	{\quotation}
	{\endquotation}

\date{}

\makeatletter
\renewcommand{\fnum@figure}{\textbf{Figure \thefigure}}
\renewcommand{\fnum@table}{\textbf{Table \thetable}}
\makeatother

\usepackage{scicite}

\usepackage{url}

\begin{document}
\def\scititle{
Strong spin–magnon coupling in a van der Waals magnet with tunable chiral symmetry
}
\title{\bfseries \boldmath \scititle}

\author
{D. Garc\'{i}a-Pons,$^{1,2\ast}$
J. P\'{e}rez-Bail\'{o}n,$^{1,2\ast}$
C. Boix-Constant,$^{3}$
I. G\'{o}mez-Muñoz,$^{3}$
\\
X. del Arco,$^{1,2}$
S. Mañas-Valero,$^{3}$
E. Coronado,$^{3\dagger}$
D. Zueco,$^{1,2\dagger}$
M. J. Mart\'{i}nez-P\'{e}rez$^{1,2\dagger}$\\
\\
\normalsize{$^{1}$Instituto de Nanociencia y Materiales de Arag\'{o}n (INMA), CSIC-Universidad de Zaragoza, Zaragoza, Spain}\\
\normalsize{$^{2}$Departamento de Física de la Materia Condensada, Universidad de Zaragoza, Zaragoza 50009, Spain}\\
\normalsize{$^{3}$Instituto de Ciencia Molecular, Universitat de Valencia, Paterna, Spain}\\
\\
\normalsize{$^\ast$These authors contributed equally to this work.}\\
\normalsize{$^\dagger$Corresponding authors: eugenio.coronado@uv.es, dzueco@unizar.es, pemar@unizar.es}
}

\maketitle
\begin{abstract}
 Quantum technologies require platforms that can coherently interface qubits with bosonic excitations. Photons have traditionally played this role in cavity quantum electrodynamics, but achieving the same goal using solid-state bosons remains challenging. Here we demonstrate strong and tunable spin–magnon coupling between molecular spin qubits and magnons in a van der Waals antiferromagnetic insulator. Using [Gd(W$_5$O$_{18}$)$_{2}$]$^{9-}$ as the spin ensemble and CrSBr as the magnonic resonator, we observe anticrossings and coherent hybridization, realizing magnon quantum electrodynamics for the first time. Crucially, by rotating the magnetic field, we can dynamically change the magnon symmetry from linear to chiral, enabling in-situ tuning of the coupling strength. Our results establish magnonic cavities as a platform for scalable chiral quantum optics with magnetic materials.
\end{abstract}


Confining photons inside cavities amplifies their coupling to matter, allowing coherent energy exchange. This is the core idea behind cavity quantum electrodynamics (QED), which has enabled decades of experiments testing the fundaments of quantum mechanics. A major breakthrough came in 2004 with the demonstration of strong coupling between a superconducting qubit and a photon in a superconducting resonator\cite{Wallraff2004}. This catalyzed the field of circuit-QED and the development of modern on-chip quantum technologies.

\begin{figure}[t]
	\includegraphics[width=0.6\columnwidth]{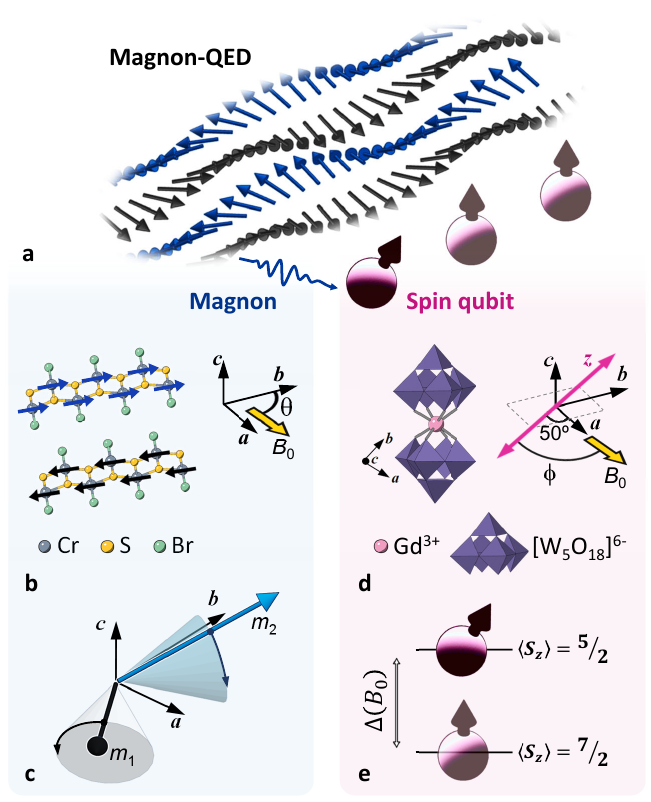}
	\caption{{\bf Magnon quantum electrodynamics.} a: One  AFM magnon induces a spin transition in a nearby spin.  b: Crystal structure of CrSBr with $\theta$ the angle between $B_0$ and the easy $b$-axis. c: Magnetization dynamics of sublattices $m_1$ and $m_2$ for the acoustic mode. d: Crystal structure of GdW$_{10}$ with $\phi$ the angle between $B_0$ and the easy $z$-axis. e: At mK temperatures, GdW$_{10}$ is an ideal two-level system with zero-field ground spin state $\langle S_z\rangle =\pm 7/2$ and first excited state $\langle S_z\rangle =\pm 5/2$ separated by $\Delta(B_0=0)=22.3$ GHz ($\sim 1.1$ K). }
	\label{fig1}
\end{figure}

Spin qubits\cite{Scappucci2020,GaitaArino2019,Doherty2013}
have so far remained outside the circuit-QED toolbox. Coupling spins  to cavities would open the door to qubit readout, the creation of spin–spin interactions\cite{Li2025,Fukami2021,Rusconi2019,Trifunovic2013},
or new phenomena like the superradiant phase transition \cite{Kim2025,RomanRoche2021}
and cavity-induced correlated states of matter\cite{Ilyas2024,Bloch2022}. Although individual spins have been addressed electrically \cite{Viennot2015,Pla2012},
strong magnetic coupling to superconducting cavities is extremely challenging. This is mainly due to a scale mismatch: microwave photons are too large to be confined within  nanoscopic resonators, needed for efficient spin–photon coupling \cite{Gimeno2020}.
Overcoming this limitation might require moving beyond circuit-QED to explore other solid-state excitations \cite{Bienfait2019}.

Spin waves are collective excitations in magnetically ordered materials, promising for  classical information processing\cite{Chumak2015}. In particular, antiferromagnets (AFMs)  are garnering growing interest\cite{Baltz2018,Jungwirth2016} due to their zero net magnetization,  high-speed dynamics and potential low-damping \cite{Das2022,Lebrun2018}.
The discovery of magnetism in van der Waals (vdW) materials, including AFM order in FePS$_3$ \cite{Lee2016}, 
has triggered interest in AFM magnonics down to the few-layer limit\cite{ManasValero2025}. In parallel, magnons are gaining attention as quantum-compatible bosonic excitations due to their nontrivial features: nonlinearity\cite{Bejarano2024}, non-reciprocity\cite{Heins2025}, quantum squeezing\cite{Yuan2022} and, in the case of AFMs, magnon bands with both chiralities\cite{Baltz2018}. These features have motivated a decade of experiments in quantum magnonics, including the coherent coupling of magnons to photons \cite{Tang2025,Zollitsch2023, MartinezPerez2018, Huebl2013},
phonons \cite{Zhang2016}, and superconducting qubits \cite{Xu2023, LachanceQuirion2019}. Yet, direct coupling to independently addressable spin qubits remains unachieved.

A natural step from cavity-QED to magnetic platforms involves replacing photons with magnons, realizing the concept of \emph{magnon–QED} (Fig.~\ref{fig1}a). A key milestone towards this goal is the observation of strong coherent coupling between magnons and solid-state spin qubits.

Spin-magnon coupling has been widely studied theoretically \cite{Li2025,Fukami2021,Rusconi2019,Trifunovic2013,Dey2025,GonzalezGutierrez2024,Candido2020,Neuman2020}.
On the experimental side, most implementations have focused on ferro- or ferrimagnetic materials but, unfortunately, have remained limited to the weak coupling regime\cite{Bejarano2024,Fukami2024}. 
AFMs, by contrast, have received less attention, since their resonances typically lie in the terahertz range  \cite{Jungwirth2016}, 
well above the energy scales relevant for most spin qubits \cite{GaitaArino2019,Doherty2013}. 
Notably, strong spin-magnon coupling has been reported in AFMs containing embedded 
paramagnetic ions, where exchange-mediated interactions serve as an effective coupling mechanism~\cite{Hiraishi2025,Li2018}. 
However, these platforms lack the key ingredients for quantum technologies: the spins are not independent qubits and the coupling cannot be externally tuned, limiting the development of scalable magnon-based quantum circuits.

Here, we experimentally demonstrate strong spin-magnon coupling between acoustic  magnonic modes in an AFM-vdW insulator and a spin ensemble, laying the groundwork for magnon-QED. 
Crucially, we show that the coupling can be tuned in situ by rotating the applied magnetic field, exploiting the intrinsic chirality of AFM magnons. This introduces a new control knob—magnon polarization symmetry—for coherent matter–matter interactions in hybrid quantum platforms.

\begin{figure*}[t]
	\includegraphics[width=0.99\textwidth]{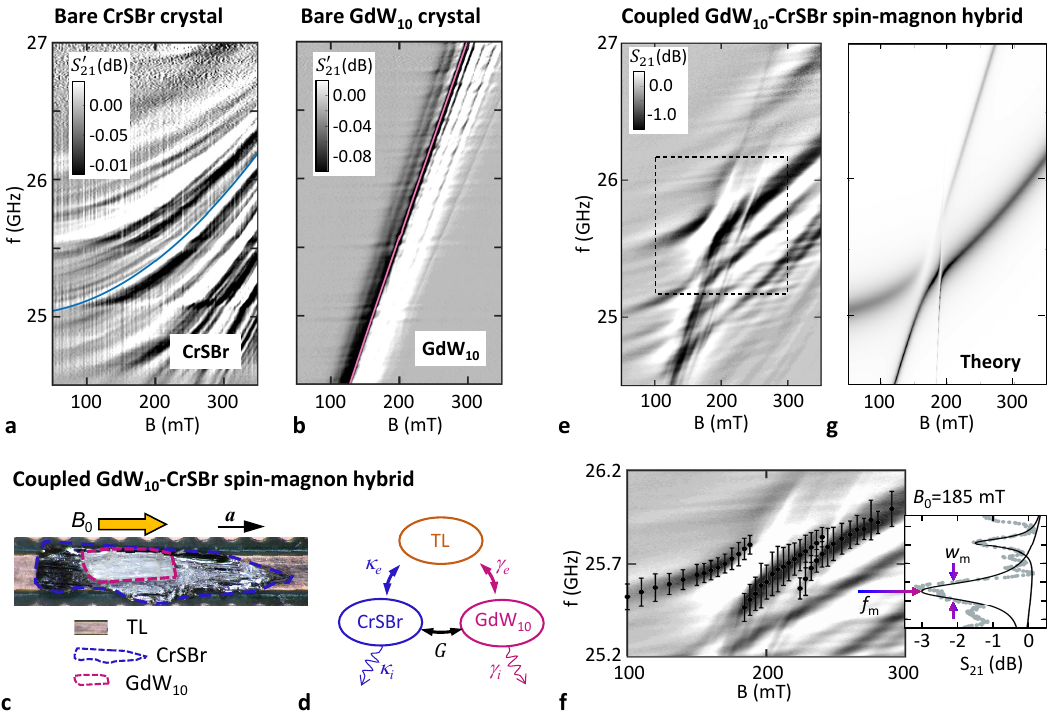}
	\caption{
    {\bf Spin-magnon hybridization in the strong coupling regime.} a (b):  Transmission pseudo-derivative spectra (see Methods) of a bare crystal of CrSBr (GdW$_{10}$) along with theoretical fitting (solid lines). 
    c: Sample assembly: GdW$_{10}$ is placed on top of CrSBr coupled to a TL. d: Relevant coupling constants and dissipation channels. e: Microwave absorption spectra of Sample1 after background subtraction (see Methods) revealing several avoided crossings indicative of strong coupling. f-right: Experimental $S_{21}$ data at $B_0 = 185$ mT (scatter points), showing a split resonance fitted with two Lorentzian curves (solid lines) with center frequency $f_m$ and linewidth $w_m$. f-left: Zoomed-in view of the dashed rectangular region in panel (e) along with fitted $f_m$ (scatter) and $w_m$ (bars).  g: Numerical simulations based on the theoretical model.
}
	\label{fig2}
\end{figure*}

\vspace{5mm}
\section{Experimental platform}

We employ CrSBr as the magnonic resonator. This layered material exhibits in-plane magnetic anisotropy, with its easy axis aligned along the crystallographic $b$-direction (Fig.~\ref{fig1}b). Below 132~K, chromium moments order ferromagnetically within each layer and antiferromagnetically between layers~\cite{Boix‐Constant2022}. When an external magnetic field $B_0$ is applied along the crystallographic $a$-axis, the magnetic sublattices undergo progressive canting, aligning with the field above $\sim 1$ T. However, all measurements in this work are performed below this critical field. In this regime, CrSBr supports two distinct spin-wave modes in the microwave range~\cite{Cham2022}: acoustic modes, where the sublattices precess in phase (Fig.~\ref{fig1}c), and optical modes, where they precess out of phase.
On the other hand, when $B_0$ is applied along the easy $b$-axis,  spin-waves acquire chiral character, corresponding to left-handed (LH) and right-handed (RH) magnon modes~\cite{Cham2022}.

To observe AFM resonance, we use a bulk CrSBr single crystal inductively coupled to a coplanar transmission line (TL). The transmitted signal ($S_{21}$ parameter) is measured at 8 mK using a vector network analyzer (VNA) and corrected as described in Methods. As shown in Fig. \ref{fig2}a, the acoustic mode splits into multiple resonances that can be precisely resolved and, notably, display narrow linewidths as low as 125 MHz (see Supplementary Information).   
The observed splitting is ascribed to staking randomness -inherent to van der Waals materials- that can be accurately described using a random stacking model (see Methods). Fig. \ref{fig2}a shows one of the resulting theoretical curves (blue line) obtained by assuming an interlayer exchange field of $\mu_0 H_{\rm E}= 0.392$~T, an in-plane anisotropy field of $\mu_0 H_{a} =0.380$~T along the medium $a$-axis, and an out-of-plane anisotropy field of $\mu_0 H_{c} =1.32$~T along the hard $c$-axis. 

As the molecular qubit, we chose the inorganic polyoxometalate [Gd(W$_5$O$_{18}$)$_{2}$]$^{9-}$, hereafter referred to as GdW$_{10}$. This single ion magnet consists of a lanthanide ion encapsulated by two [W$_5$O$_{18}$]$^{6-}$ polyanion ligands. Its sodium salt  crystallizes in a triclinic unit cell, with the easy $z$-axis in the $ab$-plane, at $\sim 50^{\circ}$ to the elongated crystal direction $a$ (Fig.~\ref{fig1}d). We have previously shown that
these lanthanoid-based molecular complexes behave as robust qubits with properties that can be tuned by an appropriate selection of the lanthanoid ion and its coordination environment \cite{Liu2021,Jenkins2017,Shiddiq2016}, and coupled to photons in superconducting cavities\cite{Carretta2021} but still far from the strong-coupling regime \cite{Gimeno2023}. In the case of GdW$_{10}$, the ground $\langle S_z \rangle = \pm 7/2$ and excited $\pm 5/2$ states are zero-field splitted by 22.3 GHz (Fig. \ref{fig1}e), effectively forming a two-level system \cite{MartinezPerez2012}. When an external magnetic field $B_0$ is applied at an angle $\phi$ to the $z$-axis, the spin states become mixtures of the $z$-projections, defining the qubit’s computational basis.

Spin transitions are characterized at 8 mK on a bare GdW$_{10}$ crystal, with $B_0$ applied along the $a$-axis (Fig. \ref{fig2}b). At low fields, the spectrum displays a single resonance line that splits into multiple resonances upon increasing $B_0$. This is attributed either to the presence of twin domains within the crystal, each with slightly different orientations of the anisotropy axis, or to local distortions in the coordination environment. The latter is likely induced by the loss of lattice water molecules during vacuum pumping used for sample cooling. In either case, the observed resonances can be well resolved at $B_0>100$ mT and exhibit  narrow linewidths, down to $30$ MHz (see Supplementary Information). Experimental data can be well accounted for by a spin Hamiltonian including zero-field anisotropy terms, $D = -1.23$ GHz and $E = 3.1$ MHz, and Zeeman energy with $\phi = 49^{\circ}$, yielding the fitted pink line in Fig. \ref{fig2}b (see Methods).

Next, we place a GdW$_{10}$ crystal on top of a CrSBr crystal with a minimum amount of apiezon-N grease, ensuring that the interaction between them is mediated by dipolar magnetic fields, with no contribution from exchange (Fig.~\ref{fig2}c). Both crystals are aligned with their elongated directions parallel to each other, such that the crystallographic $a$-axis is oriented identically in both samples. For readout, the CrSBr is placed on top of a copper TL, very much like lumped-element superconducting resonators are read out via nearby TLs in circuit-QED. Our setup enables a straightforward sample assembly and is readily adaptable to other families of vdW materials and spin qubits in solids. Here we present experimental results obtained with two samples: Sample1 shown in Fig.~\ref{fig2}c and Sample2 shown in the Supplementary Information.

\begin{figure*}[t]
	\includegraphics[width=0.7\textwidth]{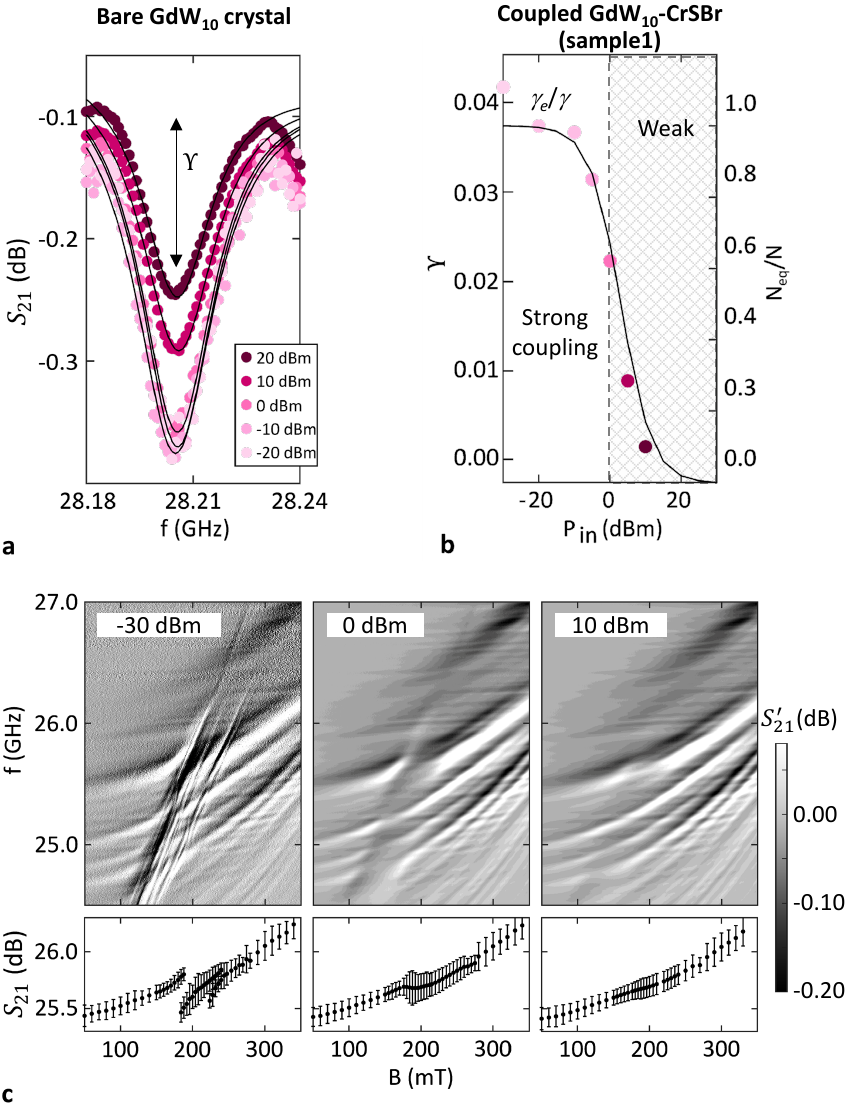}
	\caption{{\bf Power control over the spin qubit population}. a:  Transmission $S_{21}$ parameter of a bare GdW$_{10}$ crystal at different input powers given in the legend. b: Visibility $\Upsilon$  vs. power (scatter) for Sample1 at $B_0=320$ mT. Solid line shows the theoretical model. Dashed vertical line highlights the strong coupling condition $\sqrt{N_{\rm eq}/N} = \gamma/G, \kappa/G$. c: Top panels show the transmission pseudo-derivative spectra of Sample1 at three representative input powers, as indicated. Bottom panels show the results of fitting the main magnonic resonance in  transmission data (top panels) to Lorentzian curves centered at $f_m$ (dots), with linewidths $w_m$ (bars).}
	\label{fig3}
\end{figure*}

\section{Strong spin-magnon coupling}

Fig.~\ref{fig2}e shows the transmission spectra of Sample1 after background substraction. A series of clearly visible anticrossings appear when CrSBr modes intersect with GdW$_{10}$ transitions. These features are further analyzed by fitting each absorption dip to a Lorentzian function with center frequency $f_m$ and width $w_m$. At the intersection, the absorption deep splits into two clear resonances, separated by the coupling energy $2G$ (Fig.~\ref{fig2}f right). The fitted $f_m$ and $w_m$ values are superimposed on the experimental data in panel f-left, with bars corresponding to $w_m$. Here, the main anticrossing is clearly visible at $B_0=185$ mT, with a second, smaller anticrossing at $B_0=220$ mT. The fits yield a maximum experimental collective coupling strength of $G/2\pi = 130$ MHz. This is to be compared with the measured linewidth of $\kappa/2\pi = 125$ MHz for the magnonic resonator and $\gamma/2\pi = 30$ MHz for the spins (see Supplementary Information), leading a cooperativity of $c=G^2/\kappa \gamma \sim 4.5$, placing the system  within the strong coupling regime.

To confirm the interpretation of the experimental data, we employ a simplified theoretical model in which the ensemble of GdW$_{10}$ qubits couples collectively to CrSBr magnons  via a coupling strength $G$.
In addition, photons through the TL couple to the CrSBr at a rate $\kappa_e$, which, together with the intrinsic loss rate $\kappa_i$, defines the total energy decay rate of CrSBr $\kappa = \kappa_e + \kappa_i$. Similarly, the TL couples to  GdW$_{10}$ spins with rate $\gamma_e$ and intrinsic loss $\gamma_i$, yielding a total decay rate $\gamma = \gamma_e + \gamma_i$ (Fig.~\ref{fig2}d). We develop the corresponding input-output formalism to calculate the transmission spectrum $S_{21}$ (see Supplementary Information). Using experimentally determined exchange and anisotropy parameters for CrSBr and GdW$_{10}$, as well as the experimentally measured collective coupling rate $G/2\pi = 130$ MHz, we fit the model with a single free parameter, $\phi$, which represents the angle between the external magnetic field $B_0$ and the quantization axis of GdW$_{10}$. Setting $\phi = 46^{\circ}$, a value consistent with possible tilting of the GdW$_{10}$ crystal during mounting, we successfully reproduce the experimental results, as shown in Fig.~\ref{fig2}g. 

\section{Magnon-QED realization}

 The observation of an anti-crossing confirms the coupling between CrSBr and GdW$_{10}$ resonances. In the following, we demonstrate that while GdW$_{10}$ exhibits the nonlinear behavior characteristic of a collection of paramagnetic spins (qubits), the magnon modes in CrSBr retain their linear and harmonic character, even when the Gd ensemble is saturated by increasing the driving power. This establishes the CrSBr–GdW$_{10}$ system as an ideal platform for magnon–QED, where the magnons act as a linear cavity field and GdW$_{10}$ as a saturable quantum matter component. In this regime, the dynamics can be effectively captured by a Tavis–Cummings model:
\begin{equation*}
\frac{H}{\hbar}
=
\omega_M \,a^\dagger a
+
\frac{\Delta(B_0)}{2}\sum_j \sigma^z_j
+
\sum_j g_j\,\sigma^+_j a + \mathrm{h.c.},
\end{equation*}
where $\omega_M$ is the magnon frequency, $a^\dagger$ ($a$) is the magnon creation (annihilation) operator, $\Delta(B_0)$ denotes the GdW$_{10}$ transition shown in Fig.~\ref{fig2}b, $\sigma^z$, $\sigma^+$ are Pauli matrices describing this transition, $g_j$ is the single spin-single magnon coupling and $\hbar$ is the reduced Plank constant. Here, we emphasize that the collective coupling \( G \) introduced above is related to the single spin–magnon couplings via \( G = \sqrt{ \sum_j |g_j|^2} \). Note that this Hamiltonian includes only a single magnon mode—the one that couples most strongly to the spins. While additional magnon modes could in principle be included.

To demonstrate nonlinearity in the behavior of GdW$_{10}$, we examine the effect of increasing input power $P_{\rm in}$.  
Fig.~\ref{fig3}a shows the transmission through a bare GdW$_{10}$ crystal at different $P_{\rm in}$, evidencing a progressive decrease of the absorption dip for $P_{\rm in} \geq 0$~dBm. A similar trend is observed in Sample1.  To see this, we perform measurements at $B_0 > 300$ mT to ensure that the GdW$_{10}$ spins are de-coupled from CrSBr modes. In Fig.~\ref{fig3}b we plot the power dependence of the resulting visibility $\Upsilon$, defined as the amplitude of the absorption dip on a linear scale (see Fig.~\ref{fig3}a).  At very low input powers,  $\Upsilon$ tends to $\gamma_e/\gamma$. The latter quantifies the fraction of photons absorbed by the spin ensemble relative to the total number of photons. As power increases, $\Upsilon$ sharply decreases to zero.

 To describe this behavior, we solve the dynamical equations beyond the weak-driving limit, abandoning the linear response regime. 
The visibility follows $\Upsilon=\frac{\gamma_e/\gamma}{1 + 4 \Lambda^2 / (\gamma_i \gamma_{||})}$. Here, $\Lambda^2 = \alpha P_{\rm in}$, with $\alpha$  accounting for attenuation at the sample position and absorption efficiency (See Supplementary Information) and $\gamma_{||}$ being the spin relaxation rate. This model fits the data reasonably well, offering a clear physical interpretation. As the number of photons (proportional to $\Lambda^2$) approaches $\gamma_i \gamma_{||}$, the system enters a saturated regime where spins can no longer efficiently relax to equilibrium, marking the onset of non-linear transmission. Thus, $\Upsilon$ directly reflects the ratio $N_{\rm eq}/N$, i.e., the ratio of spins remaining in the ground state to total spins coupled to the TL.

On the other hand, transmission measurements of CrSBr performed at varying input powers
(Fig.~\ref{fig3}c and Supplementary Information) do not exhibit saturation effects:
the transmission response remains unchanged across the entire power range.
This confirms that the magnon dynamics in CrSBr stay within the linear-response regime,
remaining a linear harmonic oscillator
throughout the explored conditions.

\begin{figure*}[t]
	\includegraphics[width=0.99\textwidth]{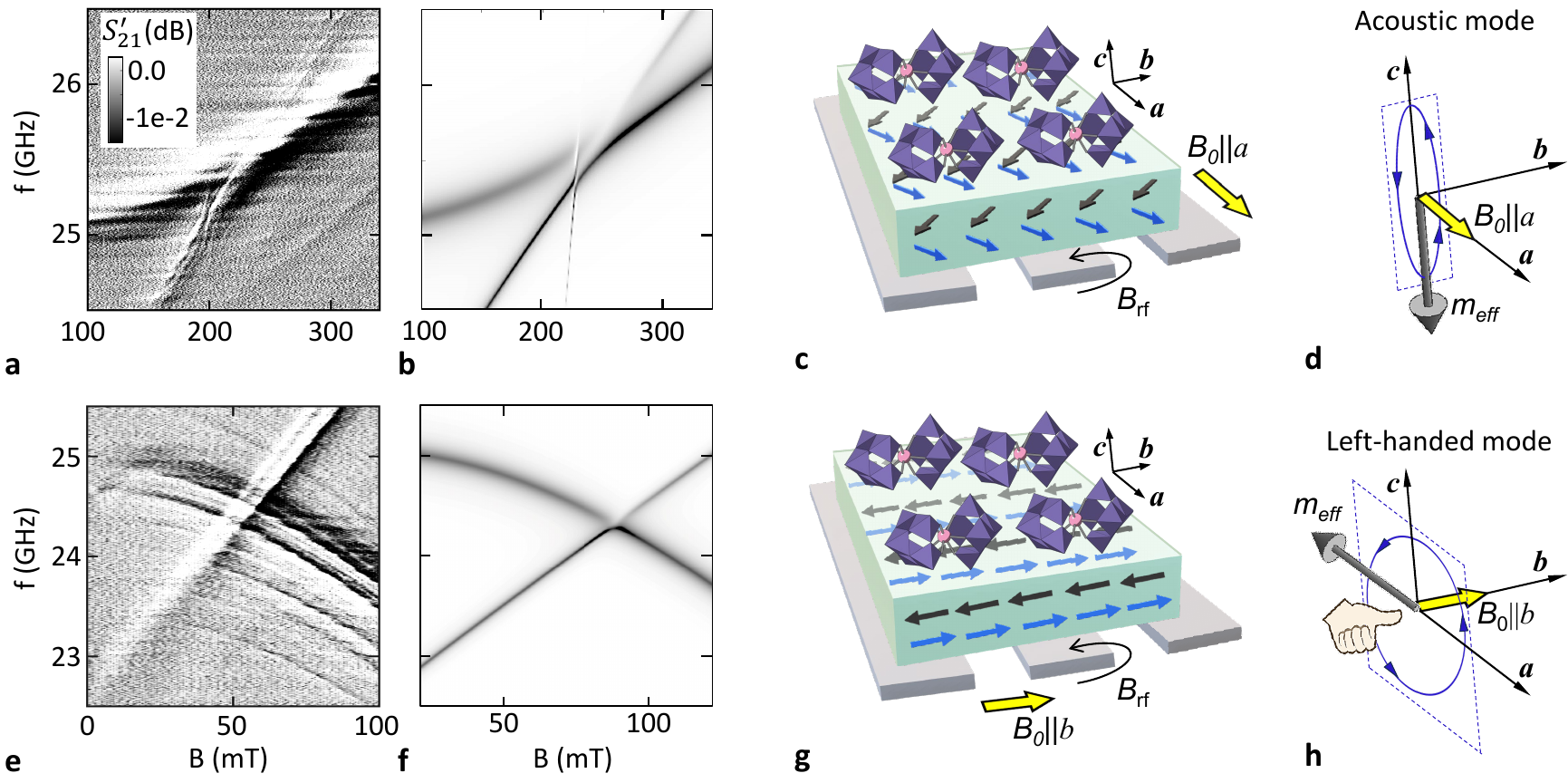}
	\caption{{\bf Dark states and coupling tuning by chiral magnons}.  a: Transmission pseudo-derivative spectra for Sample2 with $B_0$ along the $a$-axis  (acoustic mode). An avoided-crossing is visible with a bright (dark) state below (above) the anticrossing. b: Theoretical model evidencing the formation of an anti-crossing and bright/dark states.
    c: Experimental configuration for the coplanar waveguide excitation of the acoustic mode, with $B_0$ along the $a$-axis. d: Dynamics of the effective magnetization $m_{\rm eff}$ resulting from the coherent addition of both sublattice magnetizations. e: Same as in panel (a) but for $B_0$ along the $b$-axis  (LH mode). No visible signatures of coupling. f: Theoretical model corresponding to the experimental data in panel (e). g: Experimental configuration to excite the LH mode with  $B_0$ along the $b$-axis. h: Resulting $m_{\rm eff}$ precessing within the $ac$-plane, left-handed with respect to $B_0$. }
	\label{fig4}
\end{figure*}

Figure~\ref{fig3}c displays the transmission spectra of Sample1 at three $P_{\rm in}$ (top panels). The coupling stregnth is better resolved by fitting the experimental transmission data to Lorentzian curves and plotting the resulting center frequency $f_m$ and linewidht $w_m$ (bottom panels). At low power ($-30$~dBm), the primary avoided crossing is clearly visible, along with several smaller anticrossings at the intersections between CrSBr acoustic modes and GdW$_{10}$ spin transitions.  These features are reflected in the two resolved anticrossings in the bottom panel. As the drive power increases, most of the smaller anticrossings disappear, though the main one remains visible at 0~dBm. This is captured in the bottom panel: the main avoided crossing closes but the linewidths remain broadened, indicating weak coupling. For higher powers, all signatures of coupling vanish, leaving only the resonance lines associated with the CrSBr modes with no visible linewidth broadening.

This can be easily understood in view of the above discussed theory.  Reducing the effective number of equilibrium spins leads to a renormalized coupling strength, $G_{\rm eff}~=~G \sqrt{N_{\rm eq}/N}$. The strong coupling condition is then $G_{\rm eff}>\gamma, \kappa$, allowing us to calculate the threshold power for strong coupling. As shown in Fig.~\ref{fig3}b, this occurs at $P_{\rm in}=0$ dBm. Therefore, the disappearance of anticrossings results from the saturation of the GdW$_{10}$ ensemble, while nonlinearities in the spin-wave spectrum are negligible.

Next, we  we present additional evidence demonstrating that spin-magnon coupling in GdW$_{10}$-CrSBr is coherent. For this, we focus on results obtained with Sample2.
Fig.~\ref{fig4}a shows the low-power transmission at 8 mK, where a single anticrossing can be observed. The resulting coupling is slightly smaller than that of Sample1, yielding $G/2\pi = 80$ MHz. 
More importantly, these measurements evidence the presence of bright and dark states below and above the avoided-crossing, respectively. 
A similar effect is observed in Sample1, although obscured by the many anticrossings. Notice the significant reduction of the upper polaritonic dip in the right panel in Fig.~\ref{fig2}f,  indicating the formation of a dark state getting decoupled from the readout TL.

This demonstrates coherent coupling.  Consider   the single‐excitation polaritons, superpositions of the magnon mode and spin transitions,
$
|\psi\rangle = P^\dagger\,|0\rangle\otimes|S_z=-7/2\rangle,
$
with
$
P^\dagger~=~\cos\xi/2 \,a_q^\dagger + \sin\xi/2\,\frac{1}{\sqrt{7/2}\,G}\sum_j g_j S_j^+,
$
where \(|0\rangle\otimes|S_z=-7/2\rangle\) is the ground state of the CrSBr–GdW\(_{10}\) system and the mixing angle $\xi = \tan ^{-1} ( G / |\Delta(B_0) - \omega | )$  quantifies the relative spin/magnon character of the polariton.

Because both GdW\(_{10}\) and CrSBr couple to the TL, at a specific magnetic field the relative amplitudes and phases render the coupling operator $P^\dagger$ orthogonal to the polariton creation operator, decoupling the state from the TL and producing a dark state. Conversely, the other polariton branch forms a bright state with enhanced visibility (Fig.~\ref{fig2}f and ~\ref{fig4}a).

\section{Chiral-magnon tuning of the coupling}

We  exploit now the symmetry and chirality of specific magnon modes to control spin–magnon coupling. CrSBr supports LH and RH spin-wave modes, described in terms of circular
components of the magnetization precessing around the $b$-axis. In contrast, the GdW\(_{10}\) excitations are RH with respect to their local easy axis \(z\), which forms an angle \(\phi\)
with $B_0$ (see Fig.~\ref{fig1}).
As a result, the spin–magnon coupling can be selectively enhanced or suppressed
by controlling the chirality of the magnon mode and/or  the relative angle \(\phi\).

This mechanism is the basis of the emerging field of chiral  QED,
where chirality is harnessed to manipulate light–matter interactions with applications including
non-destructive discrimination and separation of enantiomers, symmetry-controlled engineering
of quantum materials, and the development of directional quantum light sources and quantum gates
\cite{Lodahl2017, Huebener2020, SuarezForero2025}. Despite this potential, implementing tunable chiral cavities remains
experimentally challenging. Here, we demonstrate that magnonic cavities, particularly  vdW AFM materials,
offer a promising platform for chiral QED due to the intrinsic handedness of  magnon modes
and the coupling mechanisms.

Experimentally, we demonstrate this effect by comparing two configurations. We focus on the behavior of the effective net magnetization, which results from the combined contribution of both magnetic sublattices. Initially, we apply the external magnetic field $B_0$ along the $a$-axis, allowing the excitation of the already discussed acoustic mode (Fig.\ref{fig4}c). Here, the effective net magnetization oscillates predominantly along the hard $c$-axis, perpendicular to the CrSBr layers (Fig.\ref{fig4}d). With the quantization (easy) axis of the GdW$_{10}$ spins oriented within the $ab$-plane, the microwave magnetic field generated by CrSBr efficiently drives spin transitions in GdW$_{10}$, resulting  in  strong spin-magnon coupling.  
Then, we rotate $B_0$ within the $ab$-plane, until it is applied parallel to the CrSBr easy $b$-axis (Fig.~\ref{fig4}g). By doing so, the resulting magnon mode progressively acquires chiral character. In particular, the acoustic mode evolves into a LH mode, in which the effective magnetization precesses 
within the \(ac\)-plane, with left-handed chirality with respect to $B_0$ (Fig.~\ref{fig4}h).  In this configuration and according to our theory, the spin-magnon coupling is expected to be reduced to approximately half of its original strength (see Supplementary Material). This residual coupling arises due to two factors: (i)  the easy $z$-axis of GdW$_{10}$ is not parallel to $B_0$, and (ii) the magnon mode is not strictly circularly polarized, but may exhibit elliptical polarization, particularly near the edges of the crystal (see Supplementary Material).

Figures \ref{fig4}a and \ref{fig4}e illustrate these contrasting scenarios for Sample2. Panel (a) corresponds to $B_0$ applied along the $a$-axis. Experiments clearly show an anticrossing, indicative of strong coupling. In contrast, panel (e) for $B_0$  along the $b$-axis reveals no coupling signatures, evidencing the selective suppression of spin–magnon interactions. Experimental results are compared to the  theoretically calculated spectra of Sample2. We use the same material parameters as for Sample1, setting $\phi$ as the only fitting parameter. Fig.~\ref{fig4}b shows the calculated absorption for the acoustic mode ($\theta=0$ and $\phi=54^{\circ}$) using $G/2\pi = 80$ MHz as determined experimentally. The model accurately reproduces the observed anticrossing, as well as the bright and dark states. In contrast,  Fig.~\ref{fig4}f presents the results obtained when exciting the LH mode ($\theta=90^{\circ}$ and $\phi=144^{\circ}$), assuming half the coupling, i.e.,  $G=40$ MHz, which is compatible with experimental measurements. 

This mechanism enables tuning the spin–magnon coupling by simply rotating the external magnetic field direction, thus switching between (quasi) linearly polarized acoustic modes to chiral LH modes within the same device. This highlights a distinct advantage of exploiting AFMs for magnon-QED, demonstrating that the coherent interaction  can be switched on or off in situ, not by redesigning the device, but simply by changing the direction of an external magnetic field.

\section{Conclusions}

To conclude, we have experimentally demonstrated that the CrSBr–GdW$_{10}$ system is accurately described by linear magnon modes and spin transitions analogous to light-matter interactions. 
These results establish  AFM-vdW materials coupled to spin qubits as a practical and tunable platform for magnon-QED, an emerging paradigm where cavity photons are replaced by discrete, externally controllable magnon modes. 
Moreover, AFM materials naturally host more than two magnetic sublattices, giving rise to several magnon branches with distinct polarization properties, including linear and chiral modes. These modes can be dynamically selected and controlled. We exploit this feature to tune the spin–magnon coupling strength within a single device—an essential capability for isolating the spin system from the magnonic resonator and thereby enhancing qubit coherence.

Our experimental results are a starting point for de systematic envelopment of 
 magnon-QED. Future studies may focus on other vdW magnets with distinct spin anisotropies, and other compatible solid-state spin systems, such as diamond color centers or silicon dopants. This might enable, for the first time, achieving magnon-mediated long-range spin–spin interactions\cite{Li2025,Fukami2021,Rusconi2019,Trifunovic2013}.
This approach offers a solution to the long-standing challenge of implementing high-fidelity two-qubit gates with spin qubits, such as NV centers.
Layered vdW magnets might also enable on-demand magnon engineering exploiting, e.g., layer twisting\cite{Chen2024,BoixConstant2023},
and optimum miniaturization of magnonic resonators down to the 2D (bilayer) limit\cite{Tang2025,Zollitsch2023}, offering a route to ultrastrong spin-magnon coupling and access to superradiant regimes\cite{Kim2025,RomanRoche2021}
and cavity material engineering\cite{Ilyas2024,Bloch2022}. 
Additionally, AFMs uniquely allow dynamic control of magnon polarization between linear and circular modes, opening new opportunities for chiral-QED\cite{Lodahl2017,SuarezForero2025,Huebener2020}.

\paragraph*{Funding:}

This work is funded and supported by the 
European Research Council (ERC) under the European Union’s Horizon 2020 research and innovation programme (948986 QFaST); 
the European Union – NextGenerationEU (Regulation EU 2020/2094), through CSIC's Quantum Technologies Platform (QTEP); 
the Arag\'{o}n Regional Government through project QMAD (E09\_23R); 
MCIN for the  Advanced Materials and the Quantum Communication programs  with funding from European Union NextGenerationEU (PRTR-C17.I1), the Generlitat Valenciana and  the Government of Aragon; 
the Spanish MCIN/AEI/10.13039/501100011033 through grants CEX2023-001286-S and CEX2024-001467-M; 
and the MCIN/AEI/10.13039/501100011033  and CEX2023-001286-S funded by MICIU/AEI /10.13039/501100011033 and the European Union FEDER through projects PID2022-140923NB-C21 and PID2023-149309OB-I00. Authors acknowledge the use of facilities and technical support from the National Facility ELECMI ICTS, node "Laboratorio de Microscopias Avanzadas (LMA)" at Universidad de Zaragoza. 
E. C. acknowledges Generalitat Valenciana  (PROMETEO Program CIPROM/2024/51) and the European Union NextGenerationEU (Cátedras Chip Program TSI-069100-2023-0012) for financial support.
D. G.-P. acknowledges a JAE-PRE grant from CSIC. 
J. P.-B. acknowledges Spanish MICIU/AEI /10.13039/501100011033 and NextGeneration EU/PRTR through grant JDC2022-049096-I.

\paragraph*{Author contributions:}
M.J. M.-P. and D. Z. conceived the project and, together with E. C., supervised the project; 
E. C., S. M.-V., C. B.-C. and I. G.-M. chosed, designed and synthesized the magnetic materials; 
M.J. M.-P., D. G.-P. and J. P.-B. built the setup, performed the measurements and analysed the data; D.Z., 
M.J. M.-P. and X. A. performed numerical simulations;
D. Z.  developed the theory; 
M.J. M.-P. and D. Z. wrote the paper with input from all authors; 

\paragraph*{Competing interests:}
There are no competing interests to declare.
\paragraph*{Data and materials availability:}
All data needed to evaluate the conclusions in the paper are present
in the paper and/or the supplementary materials. Code generated for the presented study are publicly
available in https://github.com/dzueco/spin-magnon

\clearpage 

%
\bibliographystyle{sciencemag}
\bibliography{spinmagnon}
%
%



\subsection*{Supplementary materials}
Materials and Methods\\
Supplementary Text\\
Figs. S1 to S26\\

\end{document}